\documentclass[journal]{IEEEtran}
\IEEEoverridecommandlockouts
\usepackage{cite}
\usepackage{subfigure}
\usepackage{multirow,array}
\usepackage{tikz}
\usetikzlibrary{shapes,arrows}
\usepackage{pgfplots}
\usepackage{threeparttable}
\usepackage{cite}
\pgfplotsset{compat=1.14}
\PassOptionsToPackage{hyphens}{url}\usepackage{hyperref}

\usepackage[lined,linesnumbered,ruled]{algorithm2e}
\usepackage{lscape}
\usepackage{rotating}
\usepackage{wrapfig,lipsum,booktabs}
\usepackage{subfigure}
\usepackage{color}
\usepackage[Symbol]{upgreek}
\usepackage{bm}
\usepackage{mathrsfs}

\begin{document}

\title{Recent Advances in Computer Audition for Diagnosing COVID-19: An Overview
\thanks{This work was partially supported by the JSPS Postdoctoral Fellowship for Research in Japan (ID No.\,P19081) from the Japan Society for the Promotion of Science (JSPS), Japan, and the Grants-in-Aid for Scientific Research (No.\,19F19081) from the Ministry of Education, Culture, Sports, Science and Technology (MEXT), Japan.}
}

% \author{\IEEEauthorblockN{1\textsuperscript{st} Given Name Surname}
% \IEEEauthorblockA{\textit{dept. name of organization (of Aff.)} \\
% \textit{name of organization (of Aff.)}\\
% City, Country \\
% email address or ORCID}
% \and
% \IEEEauthorblockN{2\textsuperscript{nd} Given Name Surname}
% \IEEEauthorblockA{\textit{dept. name of organization (of Aff.)} \\
% \textit{name of organization (of Aff.)}\\
% City, Country \\
% email address or ORCID}
% \and
% \IEEEauthorblockN{3\textsuperscript{rd} Given Name Surname}
% \IEEEauthorblockA{\textit{dept. name of organization (of Aff.)} \\
% \textit{name of organization (of Aff.)}\\
% City, Country \\
% email address or ORCID}
% \and
% \IEEEauthorblockN{4\textsuperscript{th} Given Name Surname}
% \IEEEauthorblockA{\textit{dept. name of organization (of Aff.)} \\
% \textit{name of organization (of Aff.)}\\
% City, Country \\
% email address or ORCID}
% \and
% \IEEEauthorblockN{5\textsuperscript{th} Given Name Surname}
% \IEEEauthorblockA{\textit{dept. name of organization (of Aff.)} \\
% \textit{name of organization (of Aff.)}\\
% City, Country \\
% email address or ORCID}
% \and
% \IEEEauthorblockN{6\textsuperscript{th} Given Name Surname}
% \IEEEauthorblockA{\textit{dept. name of organization (of Aff.)} \\
% \textit{name of organization (of Aff.)}\\
% City, Country \\
% email address or ORCID}
% }

\author{\IEEEauthorblockN{Kun Qian\IEEEauthorrefmark{1}, Bj\"orn W. Schuller\IEEEauthorrefmark{2}, and Yoshiharu Yamamoto\IEEEauthorrefmark{1} \\}
\IEEEauthorblockA{\IEEEauthorrefmark{1}Educational Physiology Laboratory, The University of Tokyo, Japan, Email:~\{qian, yamamoto\}@p.u-tokyo.ac.jp}
\IEEEauthorblockA{\IEEEauthorrefmark{2}GLAM -- Group on Language, Audio, \& Music, Imperial College London, UK,\\ Email: bjoern.schuller@imperial.ac.uk}
}

\maketitle

\begin{abstract}
Computer audition (CA) has been demonstrated to be efficient in healthcare domains for speech-affecting disorders (e.\,g., autism spectrum, depression, or Parkinson's disease) and body sound-affecting abnormalities (e.\,g., abnormal bowel sounds, heart murmurs, or snore sounds). Nevertheless, CA has been underestimated in the considered data-driven technologies for fighting the COVID-19 pandemic caused by the SARS-CoV-2 coronavirus. In this light, summarise the most recent advances in CA for COVID-19 speech and/or sound analysis. While the milestones achieved are encouraging, there are yet not any solid conclusions that can be made. This comes mostly, as data is still sparse, often not sufficiently validated and lacking in systematic comparison with related diseases that affect the respiratory system. In particular, CA-based methods cannot be a standalone screening tool for SARS-CoV-2. We hope this brief overview can provide a good guidance and attract more attention from a broader artificial intelligence community.
\end{abstract}


\section{Introduction}
\label{sec_intro}

At the time of writing, the COVID-19 pandemic (caused by the SARS-CoV-2 coronavirus) has affected more than 200 countries and regions with more than 63 million confirmed cases, leading to more than 1.4 million deaths\footnote{https://coronavirus.jhu.edu/}. To fight this global crisis, we have witnessed a plethora of endeavours made by a broad scientific community including biology, medicine, sociology, and the information communication technology (ICT). In particular, advanced technologies in the 5\,G, internet of things (IoT), and artificial intelligence (AI) have been successfully applied to the diagnosis, management, monitoring, and tracking of COVID-19. Among these AI based applications, fast screening of the chest CT has shown promising results in the current state-of-the-art works. Nevertheless, computer audition (CA) has been underestimated even though it has been increasingly studied in the domain of healthcare~\cite{qian2020computer}. 

To the best of our knowledge, the main authors of this contribution firstly presented the idea of leveraging the power of CA to combat this war between the virus and humans~\cite{schuller2020covid}. In that opinion paper, we systematically summarised the milestones achieved by advanced CA methods and their applications in speech and sound analysis, which can be possibly used in the COVID-19 context. Furthermore, we showed a preliminary experimental study on analysing the spontaneous speech sound of COVID-19 patients collected from Wuhan city, China~\cite{han2020early}. In the past eight months, we have seen an increasing number of literature focusing on CA methods for fighting the COVID-19 pandemic~\cite{brown2020exploring,imran2020ai4covid,bagad2020cough,laguarta2020covid,pinkas2020sars}. However, an overview of these existing studies is lacking, which restrains the sustainable development and guidance to the ongoing relevant works. To this end, we write this brief overview article to summarise the current achievements and indicate future directions of CA for COVID-19 studies. 

The main contributions of this work are: First, we give a concise summary of the state-of-the-art works on using CA methods for combating the COVID-19 pandemic. We investigate and compare the models by indicating their main methods and results in each literature contribution. Second, we provide an assessment to the readers to what extent CA can be exploited in practice in this global crisis given the encouraging results already achieved. On the other hand, we point out the limitations of the current studies, which should be borne on mind, and reasonably solved in future works. Third, we show  future research directions. The remainder of this paper will be laid out as follows: We firstly make a brief description of the existing databases and methods in Section~\ref{sec_dm}. Then, we present the current milestones reported in the ongoing studies in Section~\ref{sec_fingds}. Finally, we conclude this overview article in Section~\ref{sec_con}.

\section{Databases and Methods}
\label{sec_dm}

%% comparison in published literature
\begin{table*}[t]
\renewcommand{\arraystretch}{.8}
\caption{Published Literature on Computer Audition for COVID-19. MFCCs: Mel-frequency Cepstral Coefficients. GFW: Glottal Flow Waveform. SVM: Support Vector Machine. PCA: Principal Component Analysis. LR: Logistic Regression. CNN: Convolutional Neural Network. GRU-RNN: Gated Recurrent Unit based Recurrent Neural Network. DT: Decision Tree. RF: Random Forest. AB: AdaBoost. CL: Chance Level. CV: Cross Validation. LOSO-CV: Leave-One-
Subject-Out Cross Validation. \textbf{Sev.}: Severity. \textbf{S}: Sleep Quality. \textbf{F}: Fatigue. \textbf{A}: Anxiety.}
\centering
\scalebox{.85}{
\begin{threeparttable}
\begin{tabular}{llllll}
\toprule
\textbf{Ref.} & \textbf{Methods} & \textbf{\# Subjects} & \textbf{\# Instances} & \textbf{Evaluation} & \textbf{Results} \\
\midrule
\multirow{2}{*}{\cite{han2020early}} & Hand-crafted features, e.\,g., & \multirow{2}{*}{51} & \multirow{2}{*}{260} & \multirow{2}{*}{LOSO-CV} & \textbf{CL}:~33.3\,\%, UAR:~68.0\,\% for \textbf{Sev} task, 61.0\,\% \\
& MFCCs, F0, SVM as the classifier & & & & for \textbf{S} task, 46.0\,\% for \textbf{F} task, 56.0\,\% for \textbf{A} task\\
\midrule
\multirow{2}{*}{\cite{brown2020exploring}} & Hand-crafted features \& VGGish learnt  & 282, 52, & 439, 86 & User-based & \multirow{2}{*}{AUC: 0.80, 0.82, 0.80} \\
& features, LR, GBT, and SVM as the classifiers & 41 & 74 & CV & \\
\midrule
\multirow{2}{*}{\cite{imran2020ai4covid}} & MFCCs, Spectrogram,  & \multirow{2}{*}{n/a} & \multirow{2}{*}{543} & 5-fold & \multirow{2}{*}{Accuracy: 88.8\,\%} \\
& PCA, CNN, SVM &  &  & CV & \\
\midrule
\multirow{2}{*}{\cite{bagad2020cough}} & Spectrogram & \multirow{2}{*}{1\,039} & \multirow{2}{*}{3\,117} & Train/ & \multirow{2}{*}{AUC: 0.72} \\
& CNN (ResNet-18) & & & Validation & \\
\midrule
\multirow{2}{*}{\cite{laguarta2020covid}} & MFCCs & \multirow{2}{*}{5\,320} & \multirow{2}{*}{n/a} & Train/ & \multirow{2}{*}{AUC: 0.97} \\
& CNN (ResNet-50) & & & Validation & \\
\midrule
\multirow{2}{*}{\cite{pinkas2020sars}} & Attention-based Transformer, & \multirow{2}{*}{88} & \multirow{2}{*}{292} & LOSO-CV, & F1: 0.74 to 0.80, w transformer pre-training \\
& GRU-RNNs, SVM & & & 5-fold CV & F1: 0.67 to 0.70, w/o transformer pre-training \\
\midrule
\multirow{2}{*}{\cite{cmu2020covid-19-1}} & Vocal Fold Oscillation Coefficients, & \multirow{2}{*}{19} & \multirow{2}{*}{3\,835} & \multirow{2}{*}{3-Fold CV} & \multirow{2}{*}{AUC: 0.83} \\
& LR, SVM, DT, RF, AB & & &  & \\
\midrule
~\cite{cmu2020covid-19-2} & GFW, Attention-based CNN & 19 & 3\,835 & 3-Fold CV & AUC: 0.85 \\
\midrule
\multirow{2}{*}{\cite{youtube2020covid-19}} & Mel-Filter Bank Features , & \multirow{2}{*}{19} & \multirow{2}{*}{702} & \multirow{2}{*}{6-fold CV} & \multirow{2}{*}{Accuracy: 88.6\,\%, F1: 0.93} \\
& Phoneme Posteriors, SVM & & & & \\
\bottomrule
\end{tabular}
\end{threeparttable}
}
\label{tab_comparison}
\end{table*}

Table~\ref{tab_comparison} proposes a list of different methods and models used in the literature. Considering the database (including breath, cough, and speech from both healthy control and COVID-19 patients), most of the studies are still ongoing collection work. Most authors claimed in their work that the databases will be released for public research usage in the near future. 

\section{Current Findings}
\label{sec_fingds}

Generally, the current results are encouraging. However, we cannot make direct comparisons due to the different tasks, data sets, and annotation approaches. Moreover, some of the studies are lacking rigid ground-truth annotation (most of them were based on self-reported symptoms or confirmation rather than chest CT check or RT-PCR test). 

The \emph{data scarcity} is still the first challenge in the current studies. This holds in particular for highly validated data, but also for highly diverse control-group data to ensure it includes challenging other cases -- for example, other diseases affecting the respiratory system. Another general discussion is whether asymptomatic COVID-19 affected individuals can be diagnosed by audio approaches. The authors in \cite{pinkas2020sars} believe so, but this may also be a discussion of the definition of `asymptomatic'. If there are no symptoms at all, it seems unclear how the vocal production mechansisms or coughing should be affected. In addition, there is increasing evidence that COVID-19 can go without any affection of the respiratory systems, such as purely inducing diarrhea. It is unclear to which degree such cases will impact. Most importantly, we will need to gain deeper insight into features that are characteristic and the reasoning behind. Current reports are limited, for example, to description of the vocal fold oscillation patterns, or spectral changes. However, higher level impacts such as behavioural change markers should also be taken more into account. In particular, most studies so far compare different subjects for cases of COVID-19 or non-COVID-19. However, more insights will be needed comparing data from the same subjects in both states. Sensing the human mental and/or physical health status can usually not be perfectly acquired by only one modality, e.\,g., audio. Therefore, we believe by introducing more data modalities such as those accessible by wearable devices, can contribute to a higher performance of the models.

\section{Conclusion}
\label{sec_con}

In this brief overview, we proposed the achievements and limitations in the ongoing studies on CA for fighting the COVID-19. The existing literature showed encouraging results, whereas further deeply studies are still needed. As a non-invasive approach, we believe CA based models can contribute to this battle between human and virus given its many advantages such as contact-less sensing and easy to spread at low cost possible distribution as well as real-time assessment.

%% references
\bibliographystyle{IEEEtran}
\bibliography{references}

\end{document}